# MAGORINO: Magnitude-only fat fraction and $R_2$* estimation with Rician noise modelling

Computer Processing and Modelling

**Authors:**

[1,2]Timothy JP Bray MBBChir PhD

[1,3]Alan Bainbridge PhD

[4]Emma Lim

[1,3]Margaret A Hall-Craggs MBBS MD

[5]Hui Zhang PhD

[1]Centre for Medical Imaging, University College London [2]Department of Imaging, University College London Hospital [3]Department of Medical Physics, University College London Hospitals [4]Department of Imaging, Imperial College Healthcare NHS Trust, London, [5]Department of Computer Science and Centre for Medical Image Computing, University College London

**Corresponding author:**

Timothy J.P. Bray, Centre for Medical Imaging, University College London

Email: t.bray@ucl.ac.uk

## Abstract

**Purpose:** Magnitude-based fitting of chemical shift-encoded data enables proton density fat fraction (PDFF) and $R_2$* estimation where complex-based methods fail or when phase data is inaccessible or unreliable. However, traditional magnitude-based fitting algorithms do not account for Rician noise, creating a source of bias. To address these issues, we propose an algorithm for Magnitude-Only PDFF and $R_2$* estimation with Rician Noise modelling (MAGORINO).

**Methods:** Simulations of multi-echo gradient echo signal intensities are used to investigate the performance and behavior of MAGORINO over the space of clinically-plausible PDFF, $R_2$* and signal-to-noise ratio (SNR) values. Fitting performance is assessed through detailed simulation, including likelihood function visualization, and in a multi-site, multi-vendor and multi-field-strength phantom dataset and *in vivo*.

**Results:** Simulations show that Rician noise-based magnitude fitting outperforms existing Gaussian noise-based fitting and reveals two key mechanisms underpinning the observed improvement. Firstly, the likelihood functions exhibit two local optima; Rician noise modelling increases the chance the global optimum corresponds to the ground truth. Secondly, when the global optimum corresponds to ground truth for both noise models, the optimum from Rician noise modelling is closer to ground truth. Multisite phantom experiments show good agreement of MAGORINO PDFF with reference values, and *in vivo* experiments replicate the performance benefits observed in simulation.

**Conclusion:** MAGORINO reduces Rician noise-related bias in PDFF and $R_2$* estimation, thus addressing a key limitation of existing magnitude-only fitting methods. Our results offer an insight into the importance of the noise model for selecting the correct optimum when multiple plausible optima exist.

# 1 Introduction

In recent years, chemical shift-encoded MRI (CSE-MRI) has emerged as the leading method for quantifying proton density fat fraction (PDFF), an accurate and 'confounder-corrected' biomarker of tissue fat content.[1–4] PDFF measurements are now established for the assessment of hepatic steatosis,[5–7] and are also increasingly used for other applications including the pancreas, muscle and bone marrow.[8–13] CSE-MRI typically makes use of gradient echo-based acquisitions, which have the advantage that $R_2$* measurements can be extracted simultaneously, enabling quantification of iron or calcium.[12,14–16] Gradient echo-based CSE-MRI is therefore a flexible means to quantify a variety of biologic and pathological processes.

CSE-MRI relies on the fact that fat and water have different resonant frequencies, and therefore develop time-dependent phase-differences in an MRI experiment. These differences in phase can be exploited using complex-based fitting methods, which utilise the phase of the MRI signal in order to separate the fat and water signals.[1,3,17–22] Complex fitting also offers an accurate method to measure $R_2$* [23], and avoids the need for procedures such as baseline fitting, where an additional parameter is introduced into the model to capture the noise floor[24,25], or truncation, whereby data from longer echo times are discarded[26,27], which have otherwise been used to avoid noise-related bias at high $R_2$* values. However, complex signal-based fitting methods may suffer from inaccuracies in field map estimation and phase errors, may fail in areas of large $B_0$ inhomogeneity, and require phase data to be accessible and reliable. Although this may be realistic in a research setting, it can be challenging in multi-centre studies and in standard care, where sites may not all have access to expensive research agreements or dedicated software packages (this may particularly apply in countries or regions with less funding). This limits the feasibility and increases the cost of using these measurements in clinical trials and standard care.

An alternative, potentially simpler approach is magnitude-based fitting [28] [29] [30], whereby the phase is discarded. However, existing signal magnitude-based fitting fails to consider the Rician distribution of noise, leading to inaccuracies in the estimation of both PDFF and $R_2$*.

To address the vulnerability of magnitude-based fitting to Rician noise, we propose MAGORINO (Magnitude-Only fat fraction and $R_2$* estimation with Rician Noise modelling), a new fitting algorithm combining (i) Rician noise-based likelihood optimization based on the signal magnitude and (ii) two-point search of the likelihood function to ensure that both global and local optima are explored.

# 2 Theory

## 2.1 Tissue and noise models

With a gradient-echo based CSE-MRI acquisition, assuming that the fat and water signals have equal phase at $t$ = 0 (a reasonable assumption for a multi-echo gradient echo sequences), the noise-free complex signal $S$ acquired at echo time $t$ can be modelled as:

$$S(\mathrm{t} \mid \rho_{W,}\rho_f, R_2^*, f_B) = \left(\rho_W + \rho_F \sum_{m=1}^{M} r_m \exp\left(i2\pi f_{F,m} t\right)\right) \exp\left(i2\pi f_B t\right) \exp\left(-t R_2^*\right) \quad [1]$$

where $\rho_W$ and $\rho_F$ are the amplitudes of water and fat components, $f_{F,m}$ is the frequency of each spectral fat component, $r_m$ is the relative amplitude of each spectral fat component, $M$ is the total number of fat components, $f_B$ is the frequency offset due to $B_0$ inhomogeneity and $R_2^* = 1/T_2^*$ ($s^{-1}$) is an unknown relaxation constant. The standard approach is to assume that the relative amplitudes and frequency offsets of each fat component are known *a priori*; therefore the unknown parameters are $\rho_W$, $\rho_f$, $f_B$ and $R_2^*$.

With the addition of complex Gaussian noise, the measured signal $S'$ is modelled as

$$S'(t \mid \rho_{W,}\rho_f, R_2^*, f_B, \sigma^2) = S(t \mid \rho_W, \rho_f, R_2^*, f_B) + N(0, \sigma^2) + iN(0, \sigma^2) \quad [2]$$

where $N(0, \sigma^2)$ is Gaussian noise present in both real and imaginary channels and $\sigma^2$ is the noise variance.

For a single measurement, the log likelihood for the measured signal is given by

$$\log L(S', S \mid \sigma^2) = -2\log\left(\sqrt{2\pi\sigma^2}\right) - \frac{|S' - S|^2}{2\sigma^2}. \quad [3]$$

For a set of measured signals, the log likelihood becomes

$$\log L\left(\{S'_i\},\{S_i\}|\sigma^2\right) = -2n\log\left(\sqrt{2\pi\sigma^2}\right) - \sum_{i=1}^{n}\frac{|S'-S|^2}{2\sigma^2} \qquad [4]$$

where $\{S'_i\}$ is the set of measured signals, $\{S_i\}$ is the corresponding set of predicted signals based on the parameter estimates and $n$ is the number of measurements (double the number of echo times for complex data, or the number of echo times for magnitude data).

The second term is the sum of squared errors (SSE) divided by $2\sigma^2$. The maximum value for Equation [4] therefore corresponds to the minimum SSE value; meaning that the maximum likelihood estimate can be obtained minimizing the sum of squared errors (SSE), which is the widely-used nonlinear least squares approach.

Having estimated $\rho_F$ and $\rho_W$, the proton density fat fraction is calculated using

$$PDFF = \frac{\rho_F}{\rho_F + \rho_W}. \qquad [5]$$

For the signal magnitude, the noise-free signal in equation [1] becomes

$$M(\mathrm{t} \mid \rho_{W,}\rho_f, R_2^*) = \left|\rho_W + \rho_F \sum_{m=1}^{M} r_m \exp\left(i2\pi f_{F,m} t\right)\right| \exp(-tR_2^*), \qquad [6]$$

with only three unknown parameters ($\rho_W$, $\rho_f$, and $R_2^*$).

The corresponding noisy signal becomes

$$M'(t \mid \rho_{W,}\rho_f, R_2^*, \sigma^2) = M(t \mid \rho_{W,}\rho_f, R_2^*) + N_R(0,\sigma^2) \qquad [7]$$

where $N_R(0,\sigma^2)$ denotes Rician noise and 0 and $\sigma^2$ denote the mean and variance of the underlying complex Gaussian distribution. Importantly, this distribution has a nonzero mean which depends on $\sigma^2$.

Under Rician noise, the assumption of nonlinear least squares that the minimum SSE corresponds to the maximum likelihood no longer holds, and parameter estimates obtained in this way are subject to a noise-related bias arising from the nonzero mean of Rician noise.

To address this problem, parameter estimates can be obtained directly by maximizing the log likelihood.

$$\log L(\{M'_i\},\{M_i\}|\sigma^2) = \sum_{i=1}^{n}\left[\log\frac{M'_i}{\sigma^2} - \frac{\left[M'^2_i + M^2_i\right]}{2\sigma^2} + \log I_0(\frac{M'_i M_i}{\sigma^2})\right] \quad [8]$$

where $\{M'_i\}$ is the set of measured magnitude signals at different TEs, $\{M_i\}$ is the corresponding set of predicted magnitude signals, where $I_0$ is the 0$^{th}$ order modified Bessel function of the first kind.

Note that this optimization needs to be performed with sigma estimated *a priori* [31–37] – although it is possible to estimate sigma as a floating parameter in a 'one step' fitting process, this leads to underestimation in low SNR and/or high $R_2$* voxels. This underestimation arises due to overfitting - the optimizer can increase the likelihood by simultaneously reducing sigma and minimizing SSE. Underestimation of sigma effectively increases the SNR estimate towards the Gaussian regime and reduces the impact of the Rician noise model.

*A priori* sigma estimation can be achieved in multiple ways, including by acquiring multiple realisations of the same image or from the signal intensities within a homogenous region-of-interest (ROI) [31–37]. The latter approach has the advantage that an additional acquisition is not required but has the disadvantage that it can be susceptible to spatial inhomogeneity in the ROI (also note that the previous approach of estimating sigma from background is not commonly used now as it can be inaccurate with parallel imaging).

To address this susceptibility to inhomogeneity in the ROI, we propose an extension to the ROI-based method, whereby fitting with floating sigma is performed in the individual voxels within an ROI with good SNR (i.e. in the Gaussian regime, where sigma can be estimated accurately), before performing fitting with fixed sigma for the whole image. *In vivo*, muscle provides an appropriate tissue providing good SNR, low fat fraction and low $R_2$*, enabling sigma estimation. Simulations show that this approach provides similar sigma estimates to the conventional ROI-based approach but is less vulnerable to spatial inhomogeneity (see Supplementary Information Figure S1).

## 2.2 Dual optima problem and two-point search method

With magnitude-based fitting, the likelihood function has two optima: one ’true’ solution corresponding closely to the ground truth and one incorrect ’swapped’ solution with a PDFF value at the opposite end of the range (Figure 1). To ensure that both optima are explored Triay Bagur et al. developed a two-point search method whereby the fitting is initialised twice, once assuming a pure-water voxel and once assuming a pure-fat voxel, where the solution with the lower error is taken as the output for the fit.[30] However, this approach can fail in the presence of noise, since the ’true’ solution is not always globally optimal (Figure 2). Here, to increase the chance that the true solution does correspond to the global optimum, we combine this two-point search with optimisation of the Rician likelihood function in Eqn [8]. It should be noted that the two-point search method refers to exploration of the likelihood function and is distinct from the traditional use of ‘two-point Dixon’, which refers to the number of echo times.

## 2.3 Effect of parallel imaging on noise distribution

The choice of coil combination method can impact on the noise distribution. If SENSE reconstruction is used, the noise properties are equivalent to those of a spatial matched filter (SMF) reconstruction (which is the optimal coil combination with a maximised SNR of the resulting image) and a Rician distribution is expected in the magnitude data.[38] If a sum-of-squares (SoS) reconstruction is used, the distribution is noncentral chi.[38] In this case, the assumption of Rician noise is violated but remains a more appropriate approximation than the assumption of Gaussian noise. We suggest that any fitting method should be implemented with knowledge of the reconstruction method. However, an appropriate SMF-based or SENSE reconstruction should be available on most modern scanners.

# 3 Methods

## 3.1 Simulation Experiments

### 3.1.1 Fitting implementation

We implement and compare three fitting algorithms: (i) magnitude fitting with Gaussian noise model, which is an equivalent implementation of the MAGO algorithm described by Triay Bagur et al. [30], also referred to here as MAGO, (ii) magnitude fitting with Rician noise model (iii) complex fitting implementation including estimation of $f_B$.

Each method is implemented twice using two different start points (i.e. a 'two-point search' method is used); fitting is therefore run six times in total for each voxel. The initial values of $\rho_W$ (for water-dominant initialization) and $\rho_f$ (for fat-dominant initialization) are set to the maximum signal magnitude from the multiecho data, $\max_t |S_t|$, multiplied by a constant $C$. The constant $C$ compensates for reduction in the signal magnitude due to $R_2^*$ decay and chemical shift and avoids the need for empirical manual adjustment of initial values depending on scanner gain, as performed in [30]. Here, we use $C = \exp(t_{Smax} R_{2\ init}^*)$, where $t_{Smax}$ is the echo time corresponding to the maximum signal magnitude and $R_{2\ init}^*$ is the initialization $R_2^*$ value n. Specifically, for water-dominant initialization, the initial values are $\{\rho_w, \rho_f, R_2^*\} = \{C * \max_t |S_t|, 0.001, 0.1\mathrm{ms}^{-1}\}$; for fat-dominant initialization, the initial values are $\{\rho_w, \rho_f, R_2^*\} = \{0.001, C * \max_t |S_t|, 0.1\mathrm{ms}^{-1}\}$. Each of these parameters are assigned a lower bound of 0, and $R_2^*$ is assigned an upper bound of 2ms$^{-1}$. For complex fitting, for implementation $f_B$ is correctly initialized at $f_B = 0$ and is not constrained with either upper or lower bounds. All fitting is performed by maximization of likelihood functions (equivalent to minimization of error functions under Gaussian noise); this approach ensures consistency across noise models (the error function is not defined for the Rician case). For each of methods (i)-(iii), the solution providing the highest likelihood is chosen as the fit output.

The frequency offsets and relative amplitudes for the multipeak fat spectrum are matched to those used in [30] and [39], i.e. frequency shifts relative to the water peak of -3.90, -3.50, -2.70, -2.04, -0.49 and +0.50 ppm and relative amplitudes of 0.087, 0.694, 0.128, 0.004, 0.039 and 0.048.

To mimic the proposed extension for sigma estimation (detailed in Section 2.1), we first implement Rician fitting with $\sigma$ included as floating parameter in the model (maximization of equation [8]) in 100 simulated pure-water voxels with low $R_2$* (which have good SNR for each echo time, meaning that sigma can be estimated accurately) prior to fitting the full set of simulated voxels with the sigma estimated from the first step fixed. The choice of 100 voxels approximately matches the size of a typical region-of-interest (ROI) *in vivo*, and the use of pure-water, low $R_2$* voxels mimics the properties of muscle, which we use to estimate sigma *in vivo*. To compensate for slight overfitting (and underestimation of sigma), we use a simulation-derived correction factor $k$ = 1.163. Detail on the derivation of $k$ is given in Supplementary Information Figure S1. Subsequent fitting of the whole image is performed with sigma fixed to the estimate from the first step.

All fitting is performed in MATLAB 2020a (Mathworks, Natick, MA) using the *fmincon* minimizer with an interior point algorithm on a Apple iMac with 3.8GHz 8-Core Intel i7 processor. The processing time for both MAGORINO and MAGO was approximately 0.01s per voxel, and approximately 20 minutes for a 320x320 image slice.

### 3.1.2 Experiment design

To determine the effect of varying PDFF and $R_2$* on parameter estimation, simulations were performed across a dense grid of PDFF and $R_2$* combinations, with PDFF values between 0% and 100% (at 2% intervals) and $R_2$* values between 0 and 1 $ms^{-1}$ (at 0.1$ms^{-1}$ intervals). For each PDFF/R2* combination, 1000 noise-free signals were simulated using Equation [1] and sampled at echo times corresponding to a typical *in vivo* protocol at 3T using the shortest available echo times ($TE_1$=1.1ms and ΔTE=1.1ms) [30]. Gaussian noise was added to the noise-free signals in both the real and imaginary channels according to the SNR. The simulations were performed at 'typical SNR' for a 3 Tesla protocol *in vivo* (SNR=60)[30] and at 'low SNR' (SNR = 20). Two-point Gaussian magnitude fitting (MAGO), Rician magnitude fitting (MAGORINO) and complex-fitting were then applied to the noisy signals to obtain PDFF and $R_2$* estimates.

Algorithm performance was assessed in three domains: (a) parameter error, specifically the mean error on PDFF, $R_2$* and $S_0$ estimates, where $S_0 = \rho_f + \rho_w$, (b) parameter standard deviation for PDFF, $R_2$* and $S_0$, and (c) fitting error, assessed in terms of (i) the sum-of-squared error (SSE), (ii) the sum-of-squared error relative to the noiseless signal generated directly from the ground truth parameter values in the simulation, referred to here as the 'true SSE', and (iii) the estimated SSE of the noise compared to the true noise SSE. Note that (ii) and (iii) inform on the degree of overfitting, which results in an increase in true SSE and a reduction in the estimated noise. Note also that high performance in PDFF estimation should produce both low parameter error and low standard deviation, and that in some cases consistently poor performance (consistent fat-water swaps) can produce low standard deviation.

### 3.1.3 Interrogation of specific PDFF/R2* combinations

To gain further insights into the differences in behavior between Gaussian and Rician fitting, we selected specific PDFF/$R_2$* combinations showing larger error for more detailed analysis. Specifically, for the chosen PDFF/$R_2$* combinations, the parameter estimates from fat- and water-dominant initializations were displayed for each simulation instantiation using (i) fit success histograms and (ii) likelihood difference plots. The fit success histograms show the frequency of parameter estimates over all simulation instantiations, displayed on a histogram relative to the ground truth. The likelihood difference plots are scatterplots in which the parameter estimates are plotted against the difference in likelihood for the fat-dominant and water-dominant initializations. Assuming that the initialization functions correctly (and both fat-dominant and water-dominant optima are obtained from the fits), there are two main possibilities which can be captured by the likelihood difference plot: (i) the water-dominant solution is more likely than the fat-dominant solution, resulting in a positive likelihood difference for a low PDFF estimate, (ii) the fat-dominant solution is more likely than the fat-dominant solution, resulting in a negative likelihood difference for a high PDFF estimate. Additionally, there are several further possibilities which can occur if the optimization unexpectedly finds the opposite optimum to its initialization (i.e. a fat-dominant optimization finds a water dominant-solution, or vice versa). These are (iii) both initializations find the same local optimum, resulting a likelihood difference of 0, or (iv) both initializations find the wrong local optimum, resulting in a ‘reversed’ likelihood difference such that a water-dominant solution has negative likelihood difference or a fat-dominant solution has a positive likelihood difference.

We also generated PDFF/$R_2$* scatterplots to investigate the distribution of $R_2$* values arising from the true and swapped likelihood optima.

### 3.1.4 Likelihood function visualization

To gain deeper insights into the behavior observed using fit success histograms and likelihood difference plots, we computed and visualized the likelihood functions for the chosen PDFF/$R_2$* combinations. First, noise-free data were simulated based on the PDFF/$R_2$* values chosen for interrogation, and Gaussian noise was added in real and imaginary channels. A grid of 'candidate' PDFF/R2* values was generated (PDFF 0-100% and $R_2$* 0-1$ms^{-1}$) and the likelihood at each point on the grid was computed based on Equations [3] and [8]. This 2-D likelihood plot was displayed using a colourmap, enabling identification of 'true' optima (corresponding closely to the ground truth) and swapped optima (typically with a PDFF in the opposite half of the range to the ground truth).

Having generated the 2-D likelihood plot, the noisy complex signal was passed to the fitting algorithm as above. The positions of the two candidate solutions (arising from fat-dominant and water-dominant initializations) were recorded and displayed on the likelihood function, with the chosen solution highlighted. The paths taken by the optimizer for both initializations were also displayed. A further estimate of the global optimum was obtained using a search over the generated 2-D grid of likelihood values. Note that the values from this search will generally be close to but not exactly match the values obtained from the fitting because the 2-D nature of the search means that the value of $S_0$ is fixed; this provides a useful simplification which reduces the degrees of freedom and thus reduces the potential for overfitting.

#### 3.1.5 Sigma Uncertainty

To assess the impact of inaccuracies in sigma on PDFF/$R_2$* estimation, the simulations described in Section 3.1.2 were repeated with inaccurate sigma assumptions. Specifically, the simulations were performed with a 30% overestimate and a 30% underestimate and compared against the simulations performed with accurate sigma estimates. Note that these are relatively large sigma errors that are greater than those expected based on simulation data (see Supplementary Information S1) thus providing a relatively conservative measure of the robustness of the method.

### 3.2 Phantom experiments

To ensure the feasibility of MAGORINO-based fitting on a range of scanners, MAGORINO was evaluated in a publicly-available multi-site, multi-vendor and multi-field-strength phantom dataset.[39] Full details of the phantom dataset are given in [39]. Briefly, the dataset consists of fat-water mixtures with known varying fat fraction, scanned at 1.5T and 3T at six centers (two centers each for GE, Philips and Siemens). Data acquisition was performed using each site's version of a multiecho three-dimensional spoiled gradient echo CSE sequence, with two different protocols (each performed at 1.5T and 3T) used to test the reproducibility across different acquisition parameters. Protocols 1 and 2 were performed at 1.5T and protocols 2 and 4 were performed at 3T. Protocols 1 and 3 generated approximately in-phase and opposed-phased echoes, while Protocols 2 and 4 used the shortest echoes achievable. To obtain PDFF measurements, circular ROIs were manually placed on each of the vials by a radiologist (EL). MAGORINO was implemented as described above; sigma was estimated *a priori* using fitting from the 0% PDFF vial, again using the simulation-derived correction factor $k$ = 1.163 to compensate for slight overfitting. The relationship between PDFF estimates and references values was assessed using linear regression for both MAGORINO and MAGO.

## 3.3 *In vivo* imaging

To evaluate the feasibility of MAGORINO *in vivo*, we imaged the pelvis and lower legs of two healthy volunteers. These scans were performed with institutional review board approval (Queen Square Research Ethics Committee, London, REC reference 15/LO/1475) and both subjects provided written informed consent. The data were acquired on a 3T Philips Ingenia system using a multiecho three-dimensional spoiled gradient echo sequence with monopolar readouts and flyback gradients ($TE_1$ 1.2ms, ΔTE 1.6ms, flip angle 5°, TR 25ms, matrix size 320x320, pixel spacing 1.8 x 1.8mm). Coil combination was performed using SENSE (factor 1). MAGORINO was implemented as described above; sigma was estimated from homogenous regions of muscle close to the tissues of interest. To provide a qualitative performance assessment, the images were assessed and compared by a consultant radiologist with 9 years of MRI research experience. To provide a quantitative performance assessment, particularly focusing on high $R_2$* tissues (where performance is expected to diverge from MAGO), we evaluated differences between parameter estimates in bone marrow and cortex, which are normal tissues that have high $R_2$*. To enable a direct comparison between the *in vivo* data and the predictions of simulation, simulations were re-run using the same PDFF and R2* values observed in the *in vivo* data. Bias was determined for Gaussian fitting relative to Rician fitting for both *in vivo* and simulation data.

# 4 Results

## 4.1 Simulations

Results of the simulation experiments are shown in Figure 3 and Supplementary Figures 2,3 and 4. Figure 3 shows the parameter error, Supplementary figure 2 shows the parameter standard deviation and Supplementary figure 3 shows the fitting error on PDFF, $R_2$* and $S_0$ at good SNR. Supplementary Figure 4 shows the parameter error for low SNR. The subsequent analysis for specific 'interrogated' PDFF/R2* combinations, using fit success histograms, likelihood difference plots and likelihood function visualization, is shown in Figures 4-7.

### 4.1.1 Parameter error

Figure 3 shows the mean error on PDFF, $R_2$* and $S_0$ relative to the ground truth values. Note that, for fat fraction estimates (top row), areas of positive error at low fat fraction and negative error at high fat fraction arise predominantly from fat-water swaps.

For PDFF measurements (top row), at low $R_2$* (left edge of plots), all three algorithms show minimal bias over the full range of PDFF values. For higher $R_2$* values, Gaussian fitting shows a substantial positive bias in PDFF measurements for PDFF values < 50% with $R_2$* > 0.4 (see top right quadrant of 3a), indicating frequent fat-water swaps. Rician and complex fitting both substantially reduce this bias (upper right quadrant of 3b). A caveat is that, for Rician fitting, there is a small increase in bias at high fat fraction and high $R_2$* (bottom right quadrant of 3b). Complex fitting with fB fixed (3d) almost eliminates the bias observed with the first three methods except for $R_2$* values close to 1ms$^{-1}$.

For $R_2$* measurements (middle row), Gaussian fitting carries a substantial negative bias in $R_2$* measurements, which is most severe at high $R_2$* and in the intermediate PDFF range. This bias is substantially reduced for by Rician and complex fitting, with similar performance for both algorithms.

For $S_0$ measurements (bottom row), the results largely mirror those of R2* error: Gaussian fitting carries a negative bias at high R2* measurements, and the bias is reduced for Rician and complex fitting.

The benefits of Rician fitting over Gaussian fitting in terms of reduced bias are even more pronounced at low SNR, with substantial reductions in bias for PDFF and R2* estimation (Supplementary Figure S4).

#### 4.1.2 Parameter standard deviation

Supplementary Figure S2 shows the standard deviation of PDFF, $R_2$* and $S_0$ estimates.

For PDFF measurements (top row), as $R_2$* increases, all three algorithms show an increase in PDFF SD. Note that Gaussian fitting shows areas of low PDFF SD in areas of frequent swapping (i.e. the areas of high bias in Figure 3a), whereas there are no corresponding areas of low PDFF SD associated with high bias for either Rician or complex fitting. The PDFF SD is markedly reduced for complex fitting with fixed fB.

For both $R_2$* and $S_0$, parameter variance increases with increasing $R_2$* and is broadly similar between algorithms.

#### 4.1.3 Fitting error

Supplementary Figure S3 shows the SSE (top row), 'true SSE' (i.e. SSE relative to ground truth parameter estimates) (middle row) and estimated noise SD relative to the true noise SD (bottom row).

For Gaussian fitting, there is a substantial increase in the true SSE (e) at high $R_2$* which is not seen in the standard SSE value (a) and is accompanied by a reduction in the noise estimate relative to the true noise (i), indicating overfitting. For Rician and complex fitting, this overfitting is markedly reduced.

For complex fitting, there are areas of increased SSE, true SSE and overestimation of the noise at low $R_2$*. This could be eliminated by fixing fB, suggesting incorrect estimation of fB (despite the correct initialization) as a potential cause. The likelihood function analyses below give further insight into this behavior.

### 4.1.4 Interrogation of specific PDFF/R2* combinations

Figures 4, 5 and 6 provide further insight into the behaviour of the algorithms for three PDFF/$R_2$* combinations corresponding to the top half of the PDFF error plots shown in Figure 3 (a-c). All three combinations consisted of a PDFF of 20%; the R2* values were 0, 0.3 and 0.5$ms^{-1}$ (corresponding to the top, middle and bottom rows respectively for each Figure), which were chosen to illustrate the behaviour observed in Figure 3.

Figure 4 shows fit success histograms illustrating the frequency with which the two-point Gaussian, Rician and complex fitting algorithms finding the correct likelihood maximum (optimum) at different $R_2$* values. As $R_2$* increases, the frequency with which the swapped solution is selected increases, but this increase is mitigated by the use of Rician and complex fitting. For Gaussian fitting, at high $R_2$*, the majority of fitted solutions are incorrect (swapped) (Fig 4g) whereas the majority of solutions are correct (non-swapped) for Rician and complex fitting (Fig 4h,i).

Similarly, Figure 5 plots the difference in likelihood for the true and swapped solutions from the fitting algorithms against the chosen (higher likelihood) estimate. Again, as $R_2$* increases, the likelihood of the swapped solution arising increases, but this increase is mitigated by Rician and complex fitting. For Gaussian fitting, the incorrect (swapped) solution shows greater likelihood in a majority of simulations at high $R_2$* (Fig 5g), whereas the true (non-swapped) solution shows greater likelihood in the majority of simulations for Rician and complex fitting (Fig 5h,i).

Figure 6 shows the likelihood functions obtained for a single noise instantiation for the three chosen PDFF/$R_2$* combinations. At low $R_2$* (top and middle rows), all three methods can identify the true (non-swapped) solution. However, at higher $R_2$* (third row), for Gaussian fitting the swapped solution assumes a higher likelihood and is chosen as the fit output by the MAGO algorithm. For Rician and complex fitting, the true solution has a higher likelihood and is chosen as the fit output.

Figure 7 provides insight in to the origin of $R_2$* error observed at high R2* values, and includes fit success histograms (top row) and PDFF/$R_2$* scatterplots (bottom row). Figs 7a,d show that Gaussian $R_2$* estimates are negatively shifted relative to the ground truth value.

7d shows that this arises due to selection of the swapped optimum and also due to a downward shift in the position of both the true and swapped optimum relative to the ground truth. For Rician and complex fitting, both the number of swaps and the negative shift in the positions of the optima are reduced, contributing to a reduction in bias. Figs 7g-i show example likelihood functions for Gaussian, Rician and complex fitting at high $R_2$*, and gives further insight into the $R_2$* bias observed in Fig 3e-h and Fig7a-f. For Gaussian fitting, there are two sources of negative bias in $R_2$*. First, the position of the true optimum (closest to the ground truth) is negatively shifted relative to the ground truth, as evidenced by the position of opt1 and the MLE from grid search. Second, the fit has chosen the swapped optimum due to its higher likelihood, resulting in a further downward bias (this can be considered as an '$R_2$* swap'). For Rician and complex fitting, the local optimum is closer to the ground truth value and has also been correctly chosen as the fit output, mitigating bias and accounting for the behavior in Fig 7a-f.

### 4.1.5 Sigma Uncertainty

The accuracy of sigma estimation using three different methods is shown in Supplementary Figure S1. All three methods provide accurate sigma estimates when sigma is low (i.e. when SNR is high). However, the signal intensity-based method ('ROI sigma') is less robust to any inhomogeneity in the ROI, providing evidence for the use of the fitting-based method in this work. Note that the fitted sigma estimates slightly underestimate sigma due to a degree of overfitting, but that the use of the correction factor allowed for very accurate sigma estimation on unseen simulated data.

The impact of inaccuracies in the estimated sigma is shown in Supplementary Figure S5. Even with relatively large inaccuracies (greater than those expected from the simulations above), the performance of MAGORINO remains superior to MAGO for almost all plausible PDFF/R2* combinations (the only exception being high PDFF and high R2* voxels, which is a very rare combination *in vivo*).

## 4.2 Phantom experiments

Results of the analysis of the Hernando multisite phantom dataset (where the $R_2$* values are close to 0) are shown in Figure 8. The difference image shows excellent voxel-wise agreement between methods. Median PDFF values for all 11 phantom vials are plotted against reference fat fraction values for all sites, acquisition protocols and field strengths. Linear regression results are shown in Supplementary Table 2 (slope, intercept and R-squared) for Hernando PDFF, MAGO PDFF and MAGORINO PDFF. The results show excellent agreement between methods, with high accuracy, high linearity and small bias (reflected in R-squared coefficients close to 1, slope close to 1, intercept close to 0). As expected, the performance of MAGO and MAGORINO is very similar in this dataset, where $R_2$* is close to 0.

### 4.3 *In vivo* imaging

Images from the two subjects are shown in Figure 9 and Supplementary Figure S6, and a more detailed interrogation of the differences in parameter values between MAGO and MAGORINO is shown in Figure 10.

Both methods produced satisfactory fat-water separation across the image and good quality $R_2$* maps. In both cases, there is some 'speckle' in the images but no structured fat-water swaps. However, as predicted by the simulations, the methods diverged in regions of high $R_2$* / low SNR such as bone marrow and cortex (where proton density is lower and the presence of calcium hydroxyapatite causes rapid dephasing). The difference maps in Figure 9 and Supplementary Figure S6 and the plots in Figure 10 show that Gaussian $R_2$* measurements are negatively biased with respect to Rician $R_2$* measurements, and Figure 10 shows that the pattern of PDFF bias observed in those high $R_2$* tissues is extremely similar to the bias predicted by simulation experiments. This suggests that the simulation experiments provide an accurate prediction of performance *in vivo*.

## 5 Discussion

The choice of fitting algorithm for PDFF and $R_2$* estimation represents a tradeoff. Complex-based fitting enables resolution of fat-water ambiguity based on phase data, has a greater number of datapoints and avoids Rician noise-related parameter bias, but dictates that phase information must be accessible and reliable (an important limitation for standard care and clinical trials, where a range of scanners may be used) and can fail in areas of large $B_0$ inhomogeneity. Conversely, magnitude-based fitting can be performed without reliable phase data but suffers from bias due to the Rician nature of the noise distribution in the magnitude signal and requires fat-water ambiguity to be solved by another method. The strengths of the magnitude-based approach have led to its use as a 'final step' in processing of data from multisite studies[39] and recently motivated the development of a pure magnitude-only algorithm known as MAGO[30]. Despite producing good agreement values with complex-based fitting, this method still suffers from noise-related bias, which is an important limitation when imaging at low SNR and/or high $R_2$*. Here, we describe a new fitting algorithm known as MAGORINO that addresses a key limitation of MAGO, namely, its vulnerability to PDFF and $R_2$* bias arising from the Rician distribution of noise.

The key results of our study are as follows. Firstly, Rician noise modelling increases the chance that the global optimum corresponds to ground truth, thus increasing the chance of finding the 'true' solution and producing more accurate PDFF and $R_2$* estimates than with the erroneous Gaussian noise assumption. The advantage of MAGORINO over MAGO becomes apparent with increasing $R_2$* and/or low SNR, where the performance of MAGO begins to deteriorate but MAGORINO retains its performance. We show that this behavior arises because the difference in likelihood between the true (non-swapped) and swapped optima is, on average, greater for MAGORINO than for MAGO because of the use of Rician noise modelling. The use of the Rician noise model dictates that the true (non-swapped) optimum is selected by the algorithm in a greater proportion of cases, resulting in a reduction in PDFF bias and variance. Furthermore, our results (Figure 7) demonstrate that the true and swapped maxima typically occur at different $R_2$* values: effectively the $R_2$* measurement can also be 'swapped', further exacerbating bias. This problem is also mitigated by the MAGORINO algorithm.

Secondly, even when the global optimum corresponds to the ground truth for both noise models, the optimum from Rician noise modelling is closer to the ground truth. This reflects the ability of the Rician noise-based fitting to correctly attribute the nonzero signal intensities at longer echo times to noise rather than a reduction decay rate.

Thirdly, we demonstrate the feasibility of using MAGORINO in phantom data and *in vivo*. In the phantom data, MAGORINO showed excellent agreement with reference values and, as predicted from simulations (since $R_2$* is low in the phantoms), was equivalent to MAGO.

However, *in vivo*, at higher $R_2$* (demonstrated in regions of bone marrow and cortex), we observed bias in both $R_2$* and PDFF measurements for MAGO relative to MAGORINO, with a pattern entirely consistent with that predicted by simulation. This suggests that the simulation experiments provide an accurate indication of performance *in vivo*.

The improvement in PDFF and $R_2$* quantification provided by MAGORINO at high $R_2$* may be particularly important when imaging tissues with 'native' high $R_2$*, such as bone marrow and cortex (particularly at low field strength/SNR), and in pathological states such as iron-overload, or when using iron-based contrast agents.[12,14–16] For example, in severe iron overload, $R_2$* values are typically greater than 0.58 $ms^{-1}$ and can be as high as 2 $ms^{-1}$ [40], beyond the upper end of the range of values simulated in this study, meaning that the biases observed here are biologically and clinically relevant.

To our knowledge, this is the first study combining explicit modelling of Rician noise with magnitude-based fitting for chemical shift-encoded MRI. Previous studies have investigated the use of Rician noise and noncentral chi modelling in $R_2$* estimation in the liver [40,41] but did not consider the effect of fat, whereas Hernando et al. used Rician noise modelling to suppress chemical shift artifact due to olefinic fat suppression, but did not apply this to multiecho gradient echo imaging.[42] Triay-Bagur et al.'s MAGO algorithm described a two-point search approach to magnitude-based fitting but did not include Rician noise modelling[30]. An important contribution of our study is that combining two-point search with Rician noise modelling has a synergistic effect, particularly with regards to resolution of fat-water ambiguity.

This work has several important limitations in terms of the method itself and the validation studies. Firstly, the current implementation of MAGORINO assumes a single $R_2$* term for water and fat, whereas in some tissues the true behavior may be more complex. However, the monoexponential model is broadly accepted to be a good approximation in various tissues including liver and bone marrow. Secondly, we did not explore the effect of variations in imaging parameters such as the choice of field strength, number of TEs, acquisition geometry or volumetric imaging. As these parameters affect SNR, they are likely to impact on the success with which fat and water can be resolved. It would be desirable to evaluate the robustness of MAGORINO compared to MAGO using acquisitions with different

SNR (for example by modulating slice thickness or number of averages), but this is beyond the scope of the present study. Thirdly, even with the Rician noise model, MAGORINO cannot correctly resolve fat-water ambiguity in all voxels; this generally results in a non-structured 'speckle' across the image rather than fat-water swaps in specific image locations. Incorporation of spatial regularization may therefore be of value for improving the homogeneity of parameter estimates and would be a simple addition to the method. Fourthly, the current implementation of MAGORINO requires manual definition of a small ROI for sigma estimation, which introduces some subjectivity. Automation of this step, for example using deep learning, could help to address this issue. Finally, the phantom experiments and in vivo validation performed in this work are preliminary. In particular, the current in vivo results were obtained from healthy subjects, meaning that 'extreme' R2* values observed in states such as iron overload were not present. Despite this, our results do show an advantage of MAGORINO in normal tissues with higher R2* (particularly bone marrow). However, disease states (e.g. iron overload) or imaging with iron-based contrast agents, where MAGORINO could show a more dramatic performance benefit have not yet been studied.

## Conclusion

MAGORINO reduces Rician noise-related bias in PDFF and $R_2$* estimation, thus addressing a key limitation of traditional Gaussian noise-based magnitude-only fitting and removing a potential barrier to wider implementation of CSE-MRI in clinical care and trials.

## Acknowledgements

Timothy J.P. Bray was supported by an NIHR Clinical Lectureship (CL-2019-18-001). MHC is supported by the National Institute for Health Research (NIHR) Biomedical Research Centre (BRC). This work was undertaken at UCLH/UCL, which receives funding from the UK Department of Health's NIHR BRC funding scheme. The views expressed in this publication are those of the authors and not necessarily those of the UK Department of Health.

## Data Availability Statement

The MATLAB code used to generate the results in this paper has been preserved in it its original form, and is available at doi: 10.5281/zenodo.6483430. The latest version of the software is available at https://github.com/TJPBray/Rician-FW-MRI.

# 6 Figures

**Figure 1 – Conceptual illustration of magnitude-only resolution of fat-water ambiguity.** The Gaussian likelihood of PDFF estimates are shown for water-dominant voxels (PDFF=20%) (a,c) and fat-dominant voxels (PDFF=80%) (b,d). (a) and (b) are one-dimensional likelihood plots, whereas (c) and (d) are two-dimensional likelihood plots. The 2-D plot show the colour-coded likelihood over the clinically-feasible space of possible PDFF and $R_2$* values for a given pair of ground-truth parameter measurements. On the 1-D plots, ground truth estimates and the maximum likelihood are shown as red dotted lines and blue dotted lines respectively. On the 2D plots, the ground truth, maximum likelihood estimates from two-dimensional PDFF/$R_2$* grid search (MLE grid search) and local likelihood optima (i.e. the likelihood peak which is not globally optimum) are shown as black and white diamonds respectively. For both water-dominant and fat-dominant voxels, there are two likelihood maxima occurring at low PDFF and high PDFF; in each case the true (non-swapped) maximum can be identified on the basis that it has higher likelihood than the swapped solution, and thus the chosen solutions from the fitting correspond closely to the ground truth.

**Figure 2 – Conceptual illustration of failure in magnitude-only resolution of fat-water ambiguity in the presence of noise.** The Gaussian likelihood of fat fraction estimates for a truly water-dominant voxel (PDFF=20%) are shown in 1-D (a) and in 2-D (b). In this case, the swapped (incorrect) PDFF maximum has a higher likelihood than the true (non-swapped) solution, leading the algorithm to select the wrong solution. In the 1-D plot, this manifests as a reversal in the size of the two likelihood peaks, whereas in the 2-D plot it manifests as a swap in the positions of the MLE and local optimum (i.e. the black and white diamonds have swapped position compared to Fig 1c, above).

**Figure 3 – Parameter error is lower for Rician than for Gaussian fitting.** The plots show the colour-coded error in PDFF (a-d), $R_2$* (e-h) and $S_0$ (i-l) estimates for each combination of PDFF and $R_2$* values over all simulations, with SNR=60. For fat fraction measurements (top row), Gaussian fitting suffers from fat-water swaps as $R_2$* increases, particularly in the low PDFF range (producing the bright yellow area at the top of (a)). This problem is substantially mitigated using Rician fitting (b), which approaches the performance of complex fitting (c), albeit with a tradeoff of some increase in error at high PDFF and high $R_2$*. At low $R_2$*, both Gaussian and Rician magnitude fitting (a,b) show lower error in PDFF than complex fitting (c); the Figures below show that this is because complex fitting does not reach the true (non-swapped) likelihood maximum in every case, resulting in a small positive bias and increase parameter SD. For $R_2$* measurements (second row), Rician fitting (e) substantially reduces the negative bias occurring at high $R_2$* values for Gaussian fitting (d), with similar performance to complex fitting (f).

**Figure 4 – PDFF error arises from incorrect selection of the correct optimum – fit success histograms.** Each plot shows the frequency of fat fraction estimates relative to the ground truth value. All plots were generated with PDFF = 20%, whilst three different $R_2$* values (0, 0.3 and 0.5 $ms^{-1}$) were used to demonstrate the effect of varying $R_2$*, with each $R_2$* value on a separate row. Plots have been generated for both Gaussian fitting (a,d,g), Rician fitting (b,e,h) and complex fitting (c, f, i). For Gaussian fitting, as $R_2$* increases, the likelihood of the swapped solution arising increases, but this increase is mitigated by the use of Rician or complex fitting. For Gaussian fitting, at high $R_2$*, the majority of fitted solutions are incorrected (swapped) (e), whereas the majority of solutions are correct (non-swapped) for Rician fitting (h) and complex fitting (i).

**Figure 5 – PDFF error arises from incorrect selection of the correct optimum – likelihood difference plots.** For each plot, the y-axis shows the difference in likelihood between water-dominant and fat-dominant solutions. A positive likelihood difference indicates that the water-dominant (low PDFF) solution is more likely, whereas a negative likelihood indicates that the fat-dominant (high PDFF) solution is more likely. All plots were generated with PDFF = 20%, whilst three different $R_2$* values (0, 0.3 and 0.5 $ms^{-1}$) were used to demonstrate the effect of varying $R_2$*, with each $R_2$* value on a separate row. Plots have been generated for both Gaussian fitting (a,d,g), Rician fitting (b,e,h) and complex fitting (c, f, i). For Gaussian fitting, the incorrect (swapped) solution shows greater likelihood in a majority of simulations at high $R_2$* (g), whereas the true (non-swapped) solution shows greater likelihood in the majority of simulations for Rician fitting (h) and complex fitting (i). Note that for complex fitting can sometimes find the same minimum from both fat-dominant and water dominant initializations, resulting in a likelihood difference of zero, or find the wrong minima for both initializations, resulting in 'reversed' likelihood values (negative at low PDFF or positive at high PDFF).

**Figure 6 – PDFF error arises from incorrect selection of the correct optimum – likelihood function visualization.** To gain further insight into the behaviors observed in Figures 3-5, likelihood functions (and fitted solutions) were visualized for three R2* values (0, 0.3 and 0.5 $ms^{-1}$) with a fixed PDFF = 20%. Each plot shows the color-coded likelihood over the clinically-feasible space of possible PDFF and $R_2$* values for a given pair of ground-truth parameter measurements. Plots have been generated for both Gaussian fitting (a,d,g), Rician fitting (b,e,h) and complex fitting (c, f, i). Each plot labels the ground truth fat fraction and R2*, maximum likelihood estimate from two-dimensional PDFF/$R_2$* grid search (MLE grid search), local optimum from grid search, likelihood optima from water-dominant and fat-dominant initializations (opt1 and opt2), with the chosen solution circled as the fit output, and paths on the objective function (path1 and path2 for opt1 and opt2 respectively). Note that all three methods arrive at the true (non-swapped solution) for $R_2$*=0 and $R_2$*=0.3, but at $R_2$*=0.5 only Rician and complex fitting correctly resolve the fat-water ambiguity. Note that there is a small discrepancy between the position of the position of the MLE from the grid search and the fitting outputs; this arises because the grid search was performed in two dimensions (over PDFF and $R_2$* values, in order to match the dimensions of the likelihood plot) whereas the fitting includes $\boldsymbol{S_w}$ and $\boldsymbol{S_f}$ as separate parameters, which is more realistic but leads to a greater degree of overfitting than the idealized grid search.

For complex fitting, note that only the true (non-swapped) optimum is visible on the plots; this is because the swapped optimum occurs at a different value for $f_B$ and is therefore not observed in this 2-D grid of likelihood values which effectively has fixed $f_B$.

**Figure 7 – $R_2$* error arises from shifts in the position of the optima and incorrect selection of the correct optimum.** Fit success histograms (a-c), PDFF/R2* scatterplots (d-f) and likelihood plots (g-h). For (a-c), the y-axis shows the frequency of $R_2$* estimates relative to the ground truth value for Gaussian (a), Rician (b) and complex (c) fitting. For (d-f), the scatterplots show PDFF and $R_2$* parameter estimates. All plots were generated with PDFF = 20%, $R_2$* = 0.7 $ms^{-1}$ and SNR = 60. The histograms show a substantial downward shift away from the ground truth $R_2$* value for Gaussian fitting; for Rician and complex fitting this bias is substantially reduced and the majority of estimates cluster around the ground truth value. The scatterplots show that there are two separate reasons for the $R_2$* bias observed with Gaussian fitting: (i) fat-water swaps, with the swapped solution having lower $R_2$* than the true solution (this can be considered as an '$R_2$* swap') and (ii) a further negative bias for both true and swapped solutions relative to the ground truth and relative to solutions obtained from Rician and complex fitting.

(g-i) show likelihood plots for Gaussian magnitude, Rician magnitude and complex fitting respectively. For both Rician and complex fitting, the fit results are closer to the ground truth solution than for Gaussian fitting. The distance between the MLE and opt1 is reduced for Rician fitting compared to Gaussian fitting, indicating reduced overfitting.

**Figure 8 – Agreement of Rician (MAGORINO) and Gaussian (MAGO) PDFF with reference values on a multi-site, multi-vendor, multi-field-strength phantom dataset.** Agreement plots are shown for each of the four protocols, with individual points for each of the six sites. The black line indicates perfect agreement with reference PDFF values. 'Site 7' refers to the repeat scans at Site 1. The example images (a,b,c) are from Site 1, 3T Protocol 2, chosen to enable a direct visual comparison with [30].

**Figure 9 – Example images of the pelvis (subject 1) for Gaussian fitting (left column) and Rician fitting (middle column).** Both methods produce satisfactory fat-water separation across the image and good quality $R_2$* maps. However, as predicted by the simulations, the methods diverge in regions of high R2* / low SNR, which is particularly pronounced in the bone marrow and cortex (red arrows) as well as the skin. The difference maps (right column) show systematically higher $R_2$* values in these regions, whilst differences in PDFF can be both positive or negative. In accordance with predictions from theory and simulation, Rician-estimated $S_0$ values are higher in regions of high $R_2$* but lower in the air surrounding the patient than for Gaussian fitting.

The following figure (Fig 10) provides a more detailed graphical illustration of the differences in these Rician-regime regions within the bone marrow (the region analysed is indicated by the red 'dashed box' in c,f).

**Figure 10 – The pattern of bias for Gaussian vs Rician fitting *in vivo* (a,b) is similar to that shown by simulation (c,d) in regions where Rician noise dominates, suggesting that the simulation experiments can accurately predict performance *in vivo*.**

Parameter estimates obtained the central region of the image (indicated in Figure 9 by the dotted red line) are shown for Subject 1 as 3D scatterplots. The position of each point is dictated by the PDFF and R2* estimates from Rician fitting, and the colour of each point represents the bias (Gaussian estimate – Rician estimate).

To facilitate direct comparison with simulation, the *in vivo* values shown in (a,b) were used to simulate and fit noisy signals, thus generating the plots in (c,d). The pattern of bias is very similar for in vivo results (top row) and simulation results (bottom row) with positive bias in PDFF values less than 0.5 and $R_2$* above 0.2 for Gaussian fitting and substantial negative bias for Gaussian fitting at higher R2*. This suggests that the simulation experiments can accurately predict performance *in vivo*.

# 7 Supplementary information legends

## 7.1 Supplementary table legends

**Supplementary Information Table 1 - Summary of models and fitted parameters.** All models were initialized using both fat- and water-dominant initializations, as specified in the second column from the right. The objective functions and estimation of $\boldsymbol{f_B}$ varied between methods (i)-(iii). The constant $\boldsymbol{C}$ effectively compensates for reduction in the signal magnitude due to $\boldsymbol{R_2^*}$ decay and chemical shift and avoids the need for empirical manual adjustment of initial values depending on scanner gain, as performed in [30]. *For complex fitting, the Gaussian log likelihood is computed separately for real and imaginary channels before summation.

**Supplementary Information Table 2 - Summary of linear regression parameters for multisite phantom dataset (agreement between measured PDFF values and reference values).**

## 7.2 Supplementary figure legends

**Supplementary Information Figure S1 – Accuracy of sigma estimation methods over the range of plausible SNR values.** The figures show the estimated sigma on the y-axis against the true sigma on the x-axis over a range of values corresponding to the expected SNR at typical clinical field strengths *in vivo* (the SNR ranged from 20 to 70). The plots were generated by simulating and fitting 100 pure water, low $R_2$* voxels with three different degrees of inhomogeneity in the signal intensities within the ROI (a,b,c).

In (a), note that the fitted sigma estimates (red line) consistently slightly underestimate sigma due to a degree of overfitting. Note also that the degree of overfitting is very consistent, even in inhomogenous voxels. To correct for this overfitting, we used linear regression to calculate and correct the slope of the red line. Having derived this slope, we calculated a correction factor *k*, which could be applied to sigma estimates to produce the 'corrected fitted sigma' estimates. To allow a fair estimate of the performance of this correction, the slope was derived on a different set of noise instantiations to those used for testing (i.e. different 'training' and 'test' simulation data were used). The 'corrected fitted sigma' method provides a very accurate estimation on the test dataset.

In (b) and (c), the effect of inhomogeneity on sigma estimates is assessed. The inhomogeneity factor species the proportional difference between the largest and smallest $\boldsymbol{S_0}$ (where $\boldsymbol{S_0} = \boldsymbol{\rho_W} + \boldsymbol{\rho_f}$) in the voxel, with all other values evenly spaces between the largest and smallest values. In a homogenous voxel (a), all methods produce accurate sigma estimates. As the voxel becomes more inhomogenous (b, c) the fitting sigma estimates remain accurate but the ROI-based sigma estimated becomes increasingly biased. The most accurate method is the fitting-based sigma estimate with correction, referred to in the legend as 'corrected fitted sigma'.

**Supplementary Information Figure S2 - Parameter SD for SNR = 60.** The plots show the colour-coded standard deviation in PDFF (a-d), $R_2$* (e-h) and $S_0$ (i-l) estimates for each combination of PDFF and $R_2$* values over all simulations, with SNR=60. Note that parameter SD generally increased with increasing $R_2$* because fat-water swaps become more frequent (as shown in the Figure above). At low $R_2$*, both Gaussian and Rician magnitude fitting (a,b) show lower PDFF SD than complex fitting (c); the Figures below show that this is because complex fitting does not reach the true (non-swapped) likelihood maximum in every case, resulting in a small positive bias and increase parameter SD. Note that this behaviour is eliminated by fixing $\boldsymbol{f_B}$ (right hand column), although this step is likely to be unrealistic in practice.

**Supplementary Information Figure S3 - Fitting error for SNR = 60.** The plots show the grayscale-coded sum of SSE (a-d), 'true SSE' (i.e. SSE calculated relative to the ground truth) (e-h) and estimated noise (SSE / simulated noise SSE) (i-l) for each combination of PDFF and $R_2$* values over all simulations, with SNR=60. For Gaussian fitting, the 'true SSE' (e) increases substantially at higher $R_2$* values, indicating overfitting to the noise. This problem is substantially reduced by Rician magnitude fitting and complex fitting. For complex fitting (third column), SSE and noise estimates are highest at low $R_2$* values because the two-point initialization does not reach the true (non-swapped) likelihood maximum in every case. Note that this behaviour is eliminated by fixing $f_B$ (right hand column), although this step is likely to be unrealistic in practice.

**Supplementary Information Figure S4 – Parameter error for SNR = 20.** The plots show the colour-coded error in PDFF (a-d), $R_2$* (e-h) and $S_0$ (i-l) estimates for each combination of PDFF and $R_2$* values over all simulations. Note that the benefit of Rician fitting is more pronounced than for SNR = 60.

**Supplementary Information Figure S5 – MAGORINO is robust to inaccurate sigma estimation.** The plots show parameter error maps for PDFF (top row) and for R2* (bottom row) for Gaussian fitting (left column) and for Rician fitting with different sigma accuracies: the second column shows Rician fitting with sigma underestimated, the third column shows Rician fitting with correct sigma, and the right-hand column shows Rician fitting with sigma overestimated. Observe that the effect of the Rician noise model becomes more pronounced from left to right, as the estimated sigma increases (note that Gaussian fitting is equivalent to a sigma assumption of 0). If sigma is underestimated by 30% (second column), the performance becomes closer to Gaussian fitting (MAGO) (left column) than for the correct sigma estimate (third column). If sigma is overestimated by 30% (right-hand column), the differences between MAGO and MAGORINO are exaggerated. Importantly, there is very little deterioration in performance for either under- or overestimated sigma, even with the substantial sigma errors assumed here.

**Supplementary Information Figure S6 – Example images of the lower legs (subject 2) for Gaussian fitting (left column) and Rician fitting (middle column). Note that this is a challenging case where complex fitting produced a complete fat-water swap in one leg.** Both Gaussian (MAGO) and Rician (MAGORINO) fitting produce satisfactory fat-water separation across the image and good quality $R_2$* maps, albeit with some non-structured swapping in the subcutaneous fat. As with subject 1, the methods diverge in regions of high R2* / low SNR, which is particularly pronounced in the bone marrow and cortex (red arrows) as well as thie skin. The difference maps (right column) show systematically higher $R_2$* values in these regions, whilst differences in PDFF can be both positive or negative.